\documentclass[11pt]{article}
\usepackage{hyperref}
\hypersetup{
pdftitle={NDPR A Fortunate Universe},
pdfauthor={Yann Ben\'etreau-Dupin},
colorlinks=true,
linkcolor={black},
citecolor={blue},
urlcolor={blue},
}
\usepackage{hyperref,sectsty,commath,setspace,amsfonts,natbib,geometry,lastpage,fancyhdr,wrapfig,graphicx,commath,amsfonts,yfonts,array,chngpage,footmisc,relsize,amsmath,amssymb,appendix,bm}
\let\quoteOLD\quote
\def\quote{\quoteOLD\small\singlespacing}
\title{Review of Geraint F. Lewis and Luke A. Barnes,\\ \emph{A Fortunate Universe: Life in a Finely Tuned Cosmos}\footnote{Cambridge University Press, 2016, 373 pp., \$27.99, ISBN 9781107156616.}}
\date{Reviewed by Yann Ben\'etreau-Dupin, Center for Philosophy of Science, University of Pittsburgh, for \emph{Notre Dame Philosophical Reviews}\footnote{This review first appeared online at \href{http://ndpr.nd.edu/news/a-fortunate-universe-life-in-a-finely-tuned-cosmos/}{http://ndpr.nd.edu/news/a-fortunate-universe-life-in-a-finely-tuned-cosmos/}}}

\begin{document} 
 
\maketitle

This new book by cosmologists Geraint F. Lewis and Luke A. Barnes is another entry in the long list of cosmology-centered physics books intended for a large audience. While many such books aim at advancing a novel scientific theory,\footnote{See for instance books by Leonard \citet{Susskind2005}, Sean \citet{Carroll2010d}, and Lee \citet{Smolin2013}. In contrast, this book mostly pursues an explanatory goal close to that of Paul \citet{Davies2007} on the same topic.} \emph{A Fortunate Universe} has no such scientific pretense. Its goals are to assert that the universe is fine-tuned for life, to defend that this fact can reasonably motivate further scientific inquiry as to why it is so, and to show that the multiverse and intelligent design hypotheses are reasonable proposals to explain this fine-tuning. This book's potential contribution, therefore, lies in how convincingly and efficiently it can make that case.\\

There are in fact two ways to read this book. One is to see it as a response to Victor Stenger's 2011 book \emph{The Fallacy of Fine-tuning: Why the Universe is Not Designed for Us}. On this reading, \emph{A Fortunate Universe} is a popular-level adaptation of Barnes's article \citep{Barnes2012} with a similar title.\footnote{Before his passing, Stenger wrote a response to Barnes's article \citep{Stenger2012}.} A second way to read this book is as a didactic work. On this second account, the issue of the fine-tuning of the universe for life is an opportunity to survey a vast array of facts, theories, and problems in cosmology and in physics more generally. The fruitful premise that fuels the whole exposition is simple: Would the universe be different---and if so, how---if the laws and constants of physics as we know them were only slightly different? Would life be possible? What additional, missing physics---or other sorts of concept perhaps---could better explain life in the universe?\\

Indeed, at first glance, the book is structured like a long lecture: first an introductory presentation of the syllabus (Chapter 1), then the lecture properly speaking (Chapters 2 through 6), a Q\&A (Chapter 7), and finally a friendly and philosophical dialogue between the authors about more speculative matters, as would happen at the end of a meal to cap off a busy conference day (Chapter 8). It is the vivid, direct tone and writing style of a friendly physics lecture that perhaps most sets this text apart among popular-level science books about `big questions'.\\

The more philosophical chapter (Chapter 8) benefits from the dialogue format, and a more conventional style (in Chapter 5) makes the presentation of the standard picture of contemporary cosmology particularly efficient. Where the writing style is more informal, as is evident for instance in Chapter 6, the exposition is not necessarily clearer for it. At times, the casual tone is risky and misses the mark.\footnote{In order to praise the ``brilliant'' Emmy Noether, ``unsung hero of modern science,'' one can read on page 192 that she was ``an intelligent woman'', and find as supportive evidence a quote from her \emph{New York Times} obituary (by Albert Einstein), according to which she ``was the most significant creative mathematical genius thus far produced since the higher education of women began.''}\\

At its best, the presentation of physical concepts and problems is engaging and illuminating. It is accessible to lay readers in that it contains almost no mathematical formalism, yet it requires some familiarity with undergraduate-level physics. While little is done, for instance, to explicate the problem of achieving a theory of quantum gravity, the authors give an accessible introduction to string theory and a critic of its ability to solve fine-tuning problems (see pp. 302ff).\\

As is common among popular science writings, the exposition occasionally relies on the authority of the scientists invoked, loosely refers to physical laws as ``mathematical laws'' (p. 280), and is quick to judge that some philosophical discussions are ill founded. It often pushes foundational issues at the margins, but this seems to be a deliberate strategy. Lewis and Barnes play with increasingly radical `what if?' questions: they first tweak the value of some free parameters, then fundamental forces, then the second law of thermodynamic, later even the number of dimensions. With this approach, foundational and philosophical issues (about, e.g., the interpretation of probabilities or the nature of physical laws discussed in Chapters 7 and 8) often come second, as if the readers needed to be eased into them or as if they only belonged to very speculative inquiry. For all that, one may be left unsatisfied by mentions of the `entropy of the universe' when the ``famously tricky'' question of applying concepts of thermodynamics to cosmology is addressed only either as a rapid summary of Roger Penrose's Weyl Curvature Hypothesis (on p. 125 and thereafter) or as a lengthy dialogue about the arguably much less interesting `Boltzmann Brains problem' (on p. 314 and thereafter).\footnote{Interested readers may consult a rich and growing corpus in the philosophy of physics, whether on thermodynamics and cosmology \citep[see, e.g.,][]{Earman2006,Torretti2007} or on the Boltzmann Brains problem and other such skeptical arguments \citep{Winsberg2012}.}\\

Several others of the thorniest foundational issues facing contemporary cosmology that are central to the book's topic are only marginally addressed, or even left out of the book. For instance, while the authors give us the standard picture of some fine-tuning problems in cosmology (namely, the horizon problem and the flatness problem), and discuss whether inflationary cosmology solves them or not, they mostly take for granted that these are indeed problems to be solved. The authors recall Einstein's dream of a theory without free parameters,\footnote{In \citet{Schilpp1969}, p. 63, quoted on p. 294.} and candidly confess that maybe this is just asking `why?' questions like ``a four-year-old at the zoo'' (p. 323). And maybe, they argue, this is not such a bad thing since ``[s]cience has come an awfully long way with its `keep looking for answers' attitude.'' (p. 292) Nevertheless, this leaves out relevant commentary on whether the most commonly discussed instances of fine-tuning in cosmology are problematic or not \citep[see, e.g.,][]{Earman1999,McCoy2015}.\\

To be sure, this book is not and does not pretend to be an overview of foundational and philosophical issues in cosmology. Its less ambitious goal is to show in what way the universe is fine-tuned for life. This is still a very ambitious goal, as it would require defining life and its physical conditions. But ``life---even a single, `simple' cell---is a miracle of complexity.'' (p. 11) Lewis and Barnes suggest a few criteria and requirements throughout the text (such as complexity, chemistry, the use of energy, and so on), but these criteria define life almost only \emph{as we know it}. Little is said about what it could be in principle and on how much its definition should rest on strictly physical criteria; ``moral worth'', they write, is not a scientific term. ``No purpose-ometer.'' (p. 266).\\

In all the possible universes that our physical laws allow, only very few allow for complex chemistry of the kind discussed in Chapter 2; this remains true even if we look for forms of life based on other chemical elements such as silicon (pp. 267-270). Therefore, it is not clear why it cries out for an explanation that a universe with DNA-based life---or any form of chemistry-based biology---would be very rare, and yet that we should find ourselves located in one of these rarities.\footnote{Lewis and Barnes repeatedly dismiss the explanatory role of selection effects: ``selection effects, by themselves, explain nothing'' (p. 277). It is unclear why they would make that claim since selection effects would only be invoked in the context of some of the physics discussed through Chapters 2-6 (and therefore not just ``by themselves''). Still, they do not see it as being explanatory enough: ``The answer to the question `why are galaxies so bright?' is not `because otherwise we wouldn't see them?''' (p. 277). Their explanatory demands in the face of such broad `why?' questions will not be satisfied until they have some deeper physics generating those phenomena (whether galaxies or universes). After all, it is the job of cosmologists to explore such explanations.}\\

A footnote (on p. 242) swiftly rules out a proposal by Anthony \citet{Aguirre2001} according to which a cold Big Bang could be life permitting on the basis that it involves ``physics that is largely unknown.'' This is surprisingly quick since Chapter 5 spent a good amount of time arguing that much of our standard model of the universe rests on physics yet unexplained. The authors justify this ruling by later arguing (on p. 288) that if a life-permitting universe is rare enough among all the possible universes according to the physics that we know, then we should not expect it to be more common based on physics we do not know. But as they acknowledge in that passage, the authors ``haven't made a watertight argument here (\ldots).''\\

This book may be of interest to philosophers precisely in virtue of these shortcomings, not in spite of them. It provides a big picture of the physics of fine-tuning, mostly accessible in lay terms, and gives aspiring philosophers of physics a taste of the tone and intellectual style one can find at cosmology conferences. Beyond that, it does so by showing the readers that a response from philosophers might be welcome. Because the authors make clear how their thinking is informed by works in metaphysics, philosophy of physics, epistemology, and the philosophy of religion, they tell the readers how they think philosophy does or could contribute, and where they think they do not know enough to see how it might.\\

One place where philosophers will surely feel pressed to react lies at the heart of Lewis and Barnes's response to the problem of fine-tuning (in Chapter 8). Lewis advocates the `multiverse' scenario: an inflaton field drives an eternal inflation, which generates a multitude of `universes', each tuned to different values for their physical parameters such as the cosmological constant; anthropic selection effects help predict where in that cosmic landscape we might find ourselves, as argued Steven \citet{Weinberg1987}. Barnes claims that a non-naturalistic explanation---namely, the intervention of a divine Creator---is more promising.
These solutions share a same argumentative structure and use of probabilities. Barnes, on p. 344, summarizes his argument for theism as follows:
\begin{enumerate}
\item Naturalism is non-informative with respect to the ultimate laws of nature.
\item Theism prefers the laws of nature that permit the existence of moral agents, such as intelligent life forms.
\item The laws and constants of nature as we know them are fine-tuned (\ldots).
\item Thus, the probability of this (kind of) universe is much greater on theism than naturalism.
\end{enumerate}
Lewis's defense of the multiverse hypothesis could be framed similarly.\footnote{To do so, substitute ``multiverse hypothesis'' for ``theism'', ``lack of explanation'' for ``naturalism'', and let step 2 be a weaker existence claim about a mechanism generating a probability distribution for different `universes'. For a more detailed criticism of such arguments, see, e.g., \citep{Norton2010,BOCP}.} Both authors agree that the set of laws referred to on steps 1 and 2 might be very different from that of step 3, but as we saw earlier, they do not claim this to be a watertight argument. That issue notwithstanding, one might argue that if as is assumed in step 1 naturalism truly is non-informative about $X$, then it ought not have anything to say about the probability of $X$. Probabilities as we usually conceive of them (whether they be `objective' or entirely epistemic in nature) do not easily allow one to express suspension of judgment or neutral support; and one cannot truly do so in a way that allows for comparative claims. Yet that is precisely what one would need for the argument to be valid.\\

Such reasoning is common in the cosmology literature, but Lewis and Barnes should be commended for laying it out in a way that will invite further conversation between physicists and philosophers.

\nocite{Stenger2011}

\bibliographystyle{apalike}

\end{document}